\documentclass[twocolumn, showpacs,preprintnumbers,amsmath,amssymb]{revtex4}

\usepackage[pdftex]{graphicx}
\usepackage{subfigure}

\begin{document}

\addtolength{\textheight}{1.2cm}
\addtolength{\topmargin}{-0.5cm}

\newcommand{\etal} {{\it et al.}}

\title{Entanglement swapping with energy-polarization entangled photons from quantum dot cascade decay}

\author{F. Troiani}
\affiliation{S3 Istituto Nanoscienze, Consiglio Nazionale delle Ricerche, I-41100 Modena, Italy}
\date{\today}

\begin{abstract}

We theoretically investigate the efficiency of an entanglement swapping procedure based on the use of quantum dots as sources of entangled photon pairs. 
The four-photon interference that affects such efficiency is potentially limited by the fine-structure splitting and by the time correlation between cascaded photons, which provide which-path information. 
The effect of  spectral inhomogeneity is also considered, and a possible quantum eraser experiment is discussed for the case of identical dots.

\end{abstract}

\pacs{78.67.Hc,42.50.Ex,03.67.Hk}

\maketitle

A deterministic source of entangled photon pairs represents a fundamental building block in quantum-computing and communication \cite{Pan12}. 
As the semiconductor analogue of atomic systems, quantum dots (QDs) offer interesting opportunities in this perspective. They can in fact be electrically or optically driven, so as to trigger the emission of an energy-polarization entangled photon pair, originating from the cascade decay of the biexciton state \cite{Benson00,Stevenson06,Young06,Akopian06,Hafenbrak07,Dousse10,Muller14}.
The amount of entanglement in the two-photon state can however be limited by specific features of the dot dynamics and of its optical spectrum. These include a classical uncertainty in the initial time of the decay process, dephasing of the quantum dot state, and the fine structure splitting between the excitonic transitions \cite{Troiani06,Hudson07,Pfanner08,Carmele10}. In particular, this latter feature introduces a polarization dependence in the photon energies, which tends to suppress two-photon interference and consequently degrades the emitted photons to a classically correlated pair \cite{Stace03}. Possible strategies for erasing the which-path information include spectral filtering, tuning of the fine-structure splitting \cite{Gerardot07,Mohan10} or of the cavity frequency \cite{delValle11,Schumacher12}, and time reordering of the photon pair \cite{Avron08,Troiani08}.

In order to assess the suitability of these cascaded photons for quantum-information applications, one needs however to investigate also higher-order coherences \cite{Stevenson12,Nilsson13}. In this perspective, we investigate here an entanglement swapping procedure, based on the use of two independent polarization-entangled photon pairs. The entanglement swapping allows one to entangle two photons belonging to different pairs, by performing a Bell-state measurement on their twin photons \cite{Pan12}. The efficiency of such procedure can be expressed in terms of four-photon coincidence probabilities, and is maximal only if four-photon interference is fully preserved. Here we show that the interference visibility is potentially limited not only by the fine-structure splitting of each dot, but also by the time correlation between cascaded photons. A simple manipulation of the photon polarization is also suggested, which could completely erase the which-path information in the ideal case of identical emitters. Finally, we quantify the loss of interference visibility resulting from the spectral inhomogeneity of the two dots \cite{Juska13}, and the corresponding loss in the entanglement swapping efficiency.

\begin{figure}[ptb]
\begin{center}
\includegraphics[width=8.5cm]{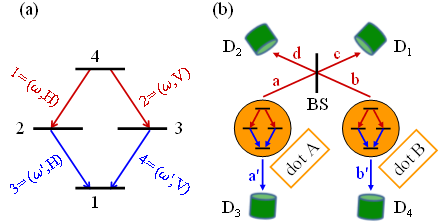}
\end{center}
\caption{(color online) (a) Level scheme of each quantum dot: $|4\rangle$ corresponds to the biexciton state, $|2\rangle$ and $|3\rangle$ to the linearly-polarized bright-exciton states, and $|1\rangle$ to the ground state. The cascade decay from state $|4\rangle$ gives rise to the sequential emission of two photons, with frequencies $\omega$ and $\omega' > \omega$, and with identical linear polarization (either $H$ or $V$). 
(b) Schematics of the experimental setup to which we refer in quantifying the efficiency of the entanglement swapping. The two red photons ($\omega$) enter the input ports of a balanced beam splitter (BS), and are detected in D$_1$ and D$_2$. The two blue photons ($\omega'$) are detected in D$_3$ and D$_4$. }
\label{fig1}
\end{figure}
{\it Dynamics of the two-photon sources --}
Each of the two-photon sources, hereafter labelled $A$ and $B$, is modelled as a four-level system (Fig. \ref{fig1}(a)). This is initially driven to state $|4\rangle$ by a delta-pulse excitation, and evolves thereafter under the effect of radiative decay ($\hbar = 1$):
\begin{equation}\label{meq}
\dot\rho = i[\rho,H_0] + 
\sum_k \frac{1}{2} \left(
2\sigma_k\rho\sigma_k^\dagger 
- \rho\sigma_k^\dagger\sigma_k 
- \sigma_k^\dagger\sigma_k\rho 
\right) ,
\end{equation}
where 
$ H_0 = 
\omega' (|2 \rangle\langle 2|+|3 \rangle\langle 3|) 
+
(\omega+\omega') |4 \rangle\langle 4| $.
The ladder operators $\sigma_k$ that enter the above master equation in the Lindblad form are given by:
\begin{equation}\label{sigmas}
\sigma_1\!\!\equiv\!\! \sqrt{\Gamma } | 2 \rangle\langle 4 |,
\sigma_2\!\!\equiv\!\! \sqrt{\Gamma } | 3 \rangle\langle 4 |,
\sigma_3\!\!\equiv\!\! \sqrt{\Gamma'} | 1 \rangle\langle 2 |,
\sigma_4\!\!\equiv\!\! \sqrt{\Gamma'} | 1 \rangle\langle 3 |.
\end{equation}
The odd (even) numbered transitions above correspond to the emission of horizontally (vertically) polarized photons.

The light emitted by each quantum dot, while radiatively relaxing to the ground state, can be characterized by second-order coherence functions of the form:
\begin{equation}
G_{ijkl} (t_1,t_3,t_4,t_2) 
= 
\langle 
\sigma_{i}^\dagger (t_1) \sigma_{j}^\dagger (t_3) 
\sigma_{k}         (t_4) \sigma_{l}         (t_2) 
\rangle ,
\end{equation}
which are computed by means of the quantum regression theorem \cite{Scully}. In particular, the relevant functions turn out to be those with $ij\!=\!13,24$ and $kl\!=\!31,42$, which for $t_1<t_2<t_3<t_4$ take the form \cite{SM}:
\begin{eqnarray}
\label{cohf1}
G_{ijkl} 
\!\!=\!\Gamma\Gamma'
e^{-2\Gamma t_1\!-(\Gamma+\Gamma'\!\!/2+i\omega)t_{21}-\Gamma' t_{32}-(\Gamma'/2+i\omega')t_{43}}\! ,
\end{eqnarray}
and vanish identically for negative $ t_{31} $ or $ t_{42} $ ($t_{ij}\equiv t_i - t_j$). 

The two-photon sources $A$ and $B$ evolve independently from one another.
Besides, for most of the present discussion, they can be assumed to be identical in terms of emission rates and frequencies. 

{\it Entanglement swapping --}
Ideally, the cascade decay of each dot $\eta$ results in the emission of two energy-polarization entangled photons:
$ |\psi_\eta^{ph}\rangle = ( | H ; H \rangle + | V ; V \rangle ) / \sqrt{2} $,
where the semicolon separates photons with frequencies $\omega$ and $\omega'$.
Being $\omega < \omega'$, we refer hereafter to these photons as the {\it red} and {\it blue} ones, respectively.
The four-photon state emitted by the two sources is thus given by:
$ |\psi_{AB}^{ph}\rangle = |\psi_A^{ph}\rangle \otimes |\psi_B^{ph}\rangle 
                         = ( | H , H ; H , H \rangle + | V , V ; V , V \rangle +
                             | H , V ; H , V \rangle + | V , H ; V , H \rangle ) / 2$.
The two blue photons are emitted independently from one another, and are in fact uncorrelated. However, they can be projected onto a maximally entangled state by means of an entanglement swapping procedure, based on the measurement of the red  photons in the Bell-states basis \cite{Pan12}. 
In particular, we consider the case where such measurement is performed within an Hong-Ou-Mandel setup (Fig. \ref{fig1}(b)), 
which allows one to distinguish the Bell state
$ | \Psi^- \rangle = (|HV\rangle - |VH\rangle) / \sqrt{2} $
from all the others.
In fact, the preparation of a state $ | \Psi^- \rangle $ at the input ports $a$ and $b$ of the beam splitter leads to the detection of coincidence events at the output ports $c$ and $d$ (coupled to the detectors $D_1$ and $D_2$, respectively), while this is not the case with any other Bell state \cite{Bouwmeester97}. 

In the Bell-state basis, the four-photon state generated by the cascade decay of the two QDs reads:  
$
|\psi_{AB}^{ph}\rangle = \frac{1}{2}
                         ( | \Phi^+_{ab} ; \Phi^+_{a'b'} \rangle 
                         + | \Phi^-_{ab} ; \Phi^-_{a'b'} \rangle
                         + | \Psi^+_{ab} ; \Psi^+_{a'b'} \rangle
                         + | \Psi^-_{ab} ; \Psi^-_{a'b'} \rangle )  
$,
where $a'$ ($b'$) is the mode corresponding to the blue photon emitted by dot $A$ ($B$), and coupled to detector $D_3$ ($D_4$).
The detection of a coincidence event by $D_1$ and $D_2$ thus singles out the last of the above components, and projects the state of the two blue photons on $ | \Psi^-_{a'b'} \rangle $. The efficiency of such entanglement swapping procedure with postselection is quantified hereafter in terms of the relevant four-photon detection probabilities.

{\it Four-photon detection probabilities --}
A state $ | \Psi^-_{ab} \rangle $ prepared at the input ports of a balanced beam splitter evolves in a state $ | \Psi^-_{cd} \rangle $ at the output ports. This state in principle gives rise to a coincidence event, corresponding to the detection of two photons with orthogonal linear polarizations at the modes $c$ and $d$. The probability that these detections take place at the times $t_1$ and $t_2$ is given by the expectation value of the operator:
\begin{eqnarray}\label{eqXcd1}
X_{cd} (t_1 , t_2 ) = \Gamma^2
\left(
a_{cH}^\dagger a_{dV}^\dagger a_{dV} a_{cH} 
+
a_{cV}^\dagger a_{dH}^\dagger a_{dH} a_{cV} 
\right) ,
\end{eqnarray}
where 
$ a_{c\chi} = a_{c\chi} (t_1) $
and
$ a_{d\chi} = a_{d\chi} (t_2) $ ($\chi = H,V$).

\begin{figure}[ptb]
\begin{center}
\includegraphics[width=8.5cm]{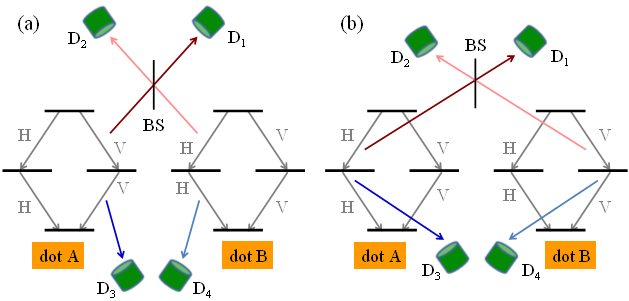}
\includegraphics[width=8.5cm]{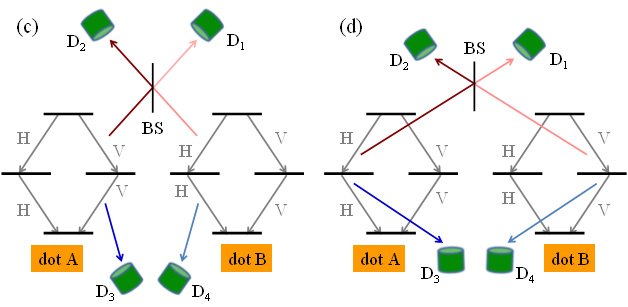}
\end{center}
\caption{(color online) All possible paths that lead to the detection of one photon by each of the four detectors, with orthogonal linear polarizations for photons of equal frequency. The two red photons, emitted by the dots $A$ and $B$, are detected in D$_1$ and D$_2$ after being both transmitted (panels (a,b)) or reflected (c,d) by the balanced beam splitter (BS). The blue photon emitted by dot $A$ ($B$) is always detected in D$_3$ (D$_4$). }
\label{fig2}
\end{figure}
In order to derive the expectation value of $ X_{cd} $, we reduce the photonic annihilation and creation operators to the ladder operators of the QDs. This is done by applying first the input-oputut relations of the 50:50 beam-splitter:
\begin{eqnarray}\label{ior}
a_{c\chi}\!\!=\!( a_{a\chi} \!\!+\!\! i a_{b\chi} ) / \sqrt{2} ,\ 
a_{d\chi}\!\!=\!( a_{b\chi} \!\!+\!\! i a_{a\chi} ) / \sqrt{2} \ (\!\chi\! =\! H,V\!) ,
\end{eqnarray}
and then the source-field relations \cite{Loudon}:
\begin{eqnarray}\label{sfr1}
a_{aH}\! =\! \sigma_{1A} ,\ 
a_{aV}\! =\! \sigma_{2A} ,\ 
a_{bH}\! =\! \sigma_{1B} ,\ 
a_{bV}\! =\! \sigma_{2B} .
\end{eqnarray}
The expression of the operator $X_{cd}$ which is obtained by combining Eqs. \ref{eqXcd1}-\ref{sfr1} reads:
\begin{eqnarray}\label{eqXcd2}
X_{cd} (t_1,t_2) \!&=&\! \frac{1}{4}
\!\sum_{k=1,2} \!\sum_{\eta=A,B}\!
\sigma_{     k      \eta }^\dagger (t_1\!)
\sigma_{\bar{k}\bar{\eta}}^\dagger (t_2\!)
\!
\nonumber\\
& & 
\left[
\sigma_{\bar{k}\bar{\eta}}         (t_2\!)
\sigma_{     k      \eta }         (t_1\!) 
- 
\sigma_{     k \bar{\eta}}         (t_2\!)
\sigma_{\bar{k}     \eta }         (t_1\!)
\right]\! ,
\end{eqnarray}
where $\bar k = 3-k$, $\bar A = B$ and $\bar B = A$. 

The time-resolved probability that the two blue photons form a Bell state can also be expressed as a combination of second-order coherence functions. In particular, if the detection of the two photons at D$_3$ and D$_4$ takes place at the times $t_3$ and $t_4$, respectively, such probability coincides with the expectation value of the operator \cite{SM}:
\begin{eqnarray}\label{Xapbp}
Y_{a'b'} (t_3 , t_4 ) & = & \frac{1}{2}
\left(
\sigma_{3A}^\dagger \sigma_{4B}^\dagger \sigma_{4B} \sigma_{3A} 
+
\sigma_{4A}^\dagger \sigma_{3B}^\dagger \sigma_{3B} \sigma_{4A} 
\right.
\nonumber\\
& - &
\left.
\sigma_{3A}^\dagger \sigma_{4B}^\dagger \sigma_{3B} \sigma_{4A}
-
\sigma_{4A}^\dagger \sigma_{3B}^\dagger \sigma_{4B} \sigma_{3A} 
\right) ,
\end{eqnarray}
where 
$ \sigma_{kA} = \sigma_{kA} (t_3) $
and
$ \sigma_{kB} = \sigma_{kB} (t_4) $ ($k=3,4$).
Here, we have made use of the source-field relations (Fig. \ref{fig1}):
\begin{eqnarray}\label{sfr2}
a_{a'H}\! =\! \sigma_{3A} ,\ 
a_{a'V}\! =\! \sigma_{4A} ,\ 
a_{b'H}\! =\! \sigma_{3B} ,\ 
a_{b'V}\! =\! \sigma_{4B} .
\end{eqnarray}

As a figure of merit of the entanglement swapping, we use the joint probability $P$ of detecting a coincidence event in the modes $c$ and $d$, and measuring the Bell state $|\Psi^-\rangle$ in $a'$ and $b'$. 
Ideally, $P$ should coincide with the weight of the state 
$ | \Psi^-_{ab} ; \Psi^-_{a'b'} \rangle $
in the four-photon state 
$ | \Psi_{AB}^{ph} \rangle $, 
which corresponds to 1/4: this value thus represents the theoretical maximum of $P$. 
In order to compute $P$, we first derive the probability density of a four-photon detection by D$_{1-4}$ at the times $t_{1-4}$, which is given by:
\begin{equation}\label{p}
p(t_1,t_2,t_3,t_4) = 
T \langle : X_{cd} (t_1,t_2) \, Y_{a'b'} (t_3,t_4) : \rangle ,
\end{equation}
where $T$ and : : indicate time and normal ordering of the ladder operators, respectively.
\begin{figure}[ptb]
\begin{center}
\includegraphics[width=8.5cm]{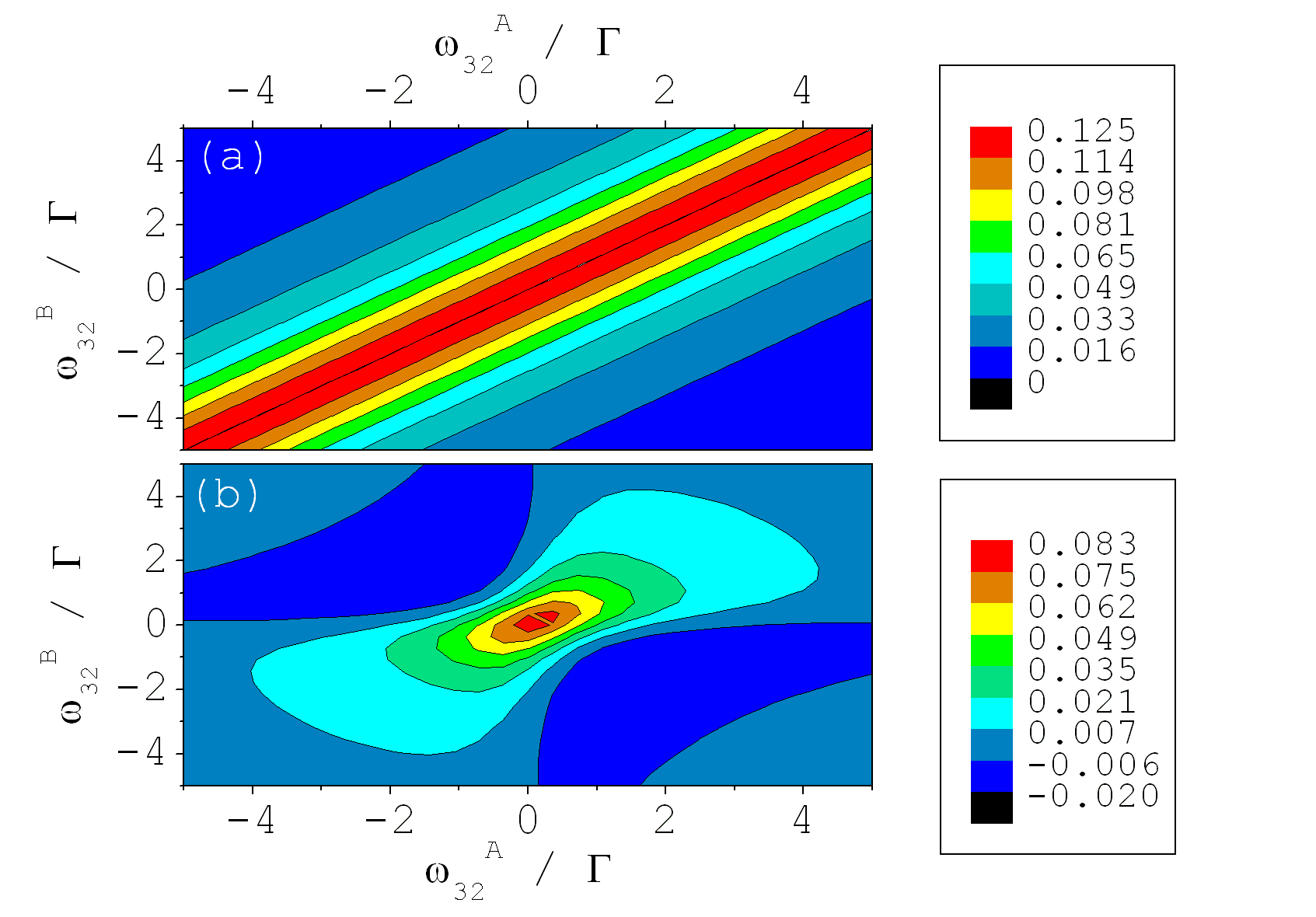}
\end{center}
\caption{(color online) Dependence of the interference term $R$ on the fine-structure splittings $\omega_{32}^\eta$ of the two dots ($\eta = A,B$). Only the intermediate energy levels are modified, according to the relations:  
$ E_{2,3}^\eta = \omega\mp\omega_{32}^\eta / 2 $. These splittings are normalized to $\Gamma = \Gamma'$.
The panels (a) and (b) refer to the cases where the partial quantum erasure is (Eq. \ref{eqR2}) and is not applied (Eq. \ref{eqR1}), respectively.}
\label{fig3}
\end{figure}
The operators $X_{cd}$ and $Y_{a'b'}$ can be replaced by the respective expressions in terms of the ladder QD operators (Eqs. \ref{eqXcd2}-\ref{Xapbp}). Being the two emitters independent from one another, each fourth-order coherence function can be factorized into the product of two second-order coherence functions (Eq. \ref{cohf1}), one for each dot. 
The time-resolved probability corresponding to the four-photon detection can be finally expressed as the sum of two kind of contributions \cite{SM}:
\begin{equation}\label{eqp}
p = \frac{1}{8} \sum_{k,l=1,2} \left( q_{kl} + r_{kl} \right) ,
\end{equation}
whose expressions in terms of the second-order coherence functions are given by: 
\begin{eqnarray}
\label{termq}
q_{kl}\!\!\! & = &\!\!
{\rm Real} \!\left[ 
G^{A}_{     l     n     n     l} (t_k       ,t_3,t_3,t_k       )\, 
G^{B}_{\bar l\bar n\bar n\bar l} (t_{\bar k},t_4,t_4,t_{\bar k}) 
\right]
\\
\label{termr}
r_{kl}\!\!\! & = &\!\!
{\rm Real} \!\left[ 
G^{A}_{     l     n\bar n\bar l} (t_k       ,t_3,t_3,t_{\bar k})\, 
G^{B}_{\bar l\bar n     n     l} (t_{\bar k},t_4,t_4,t_k       ) 
\right]
\end{eqnarray}
where 
$\bar k\!=\!3\!-\!k$, $\bar l\!=\!3\!-\!l$, $n\!=\!l\!+\!2$, $\bar n\!=\!\bar l\!+\!2$. 
Replacing the coherence functions with the expressions given in Eq. \ref{cohf1}, one can show that,
in the case of identical QDs, all the above terms in square parentheses are real and independent on the indeces $kl$:
\begin{eqnarray}\label{qrst}
q_{kl}\!\! & = &\!\!
(\Gamma\Gamma')^2 e^{(\Gamma'-2\Gamma) (t_1+t_2)-\Gamma' (t_3+t_4)}
\ (t_3\!>\!t_1, t_4\!>\!t_2) \nonumber
\\
r_{kl}\!\! & = &\!\!
(\Gamma\Gamma')^2 e^{(\Gamma'-2\Gamma) (t_1+t_2)-\Gamma' (t_3+t_4)}
\ (t_3,\!t_4 \!>\! t_1,\!t_2).
\end{eqnarray}
The overall probability $P$ is obtained by integrating $p$ from $0$ to $\infty$ with respect to the four detection times:
\begin{equation}\label{PQRST}
P \! = \! Q\! +\! R\! 
= \frac{1}{8} + \frac{1}{4} \frac{\Gamma}{2\Gamma+\Gamma'} ,
\end{equation}
where the $Q$ and $R$ come from the time integrals of all the terms $q_{kl}$ and $r_{kl}$, respectively. 
The probability $P$ thus monotonically increases from $1/8$, in the limit $\Gamma / \Gamma' \rightarrow 0 $, to the theoretical maximum $1/4$, for $ \Gamma / \Gamma' \rightarrow \infty $. 

{\it Which-path information: time correlation, spectral inhomogeneity and fine-structure splitting --}
The two contributions to the probability $P$ can be given a specific physical interpretation. In particular, the functions $q_{kl}$ are formally given by the product of probability densities (Eq. \ref{termq}), and are thus (joined) probability densities themselves. Each of them can be related to one of the possible paths that lead to the detection of one photon with frequency $\omega$ ($\omega'$) at each of the detectors D$_1$ and D$_2$ (D$_3$ and D$_4$). More specifically, $q_{kl}$ gives the time-resolved probability that photons emitted by the transitions $lA$, $nA$, $\bar lB$, $\bar nB$ are detected respectively at the times $t_k$, $t_3$, $t_{\bar k}$, $t_4$ by the detectors $D_k$, $D_3$, $D_{\bar k}$, $D_4$. The paths corresponding to $q_{12}$, $q_{11}$, $q_{22}$, and $q_{21}$ are pictorially represented in the panels (a-d) of Fig. \ref{fig2}. 

The functions $r_{kl}$ can instead be interpreted in terms of quantum interference between pairs of the above paths. In particular, the two paths corresponding to each term $r_{kl}$ (Eq. \ref{termr}) are characterized by the detection at $D_k$, $D_3$, $D_{\bar k}$, $D_4$ of photons respectively emitted by the transitions $lA$, $nA$, $\bar lB$, $\bar nB$ or by $lB$, $\bar nA$, $\bar lA$, $nB$.
They thus differ from one another both in the source of the photons detected at $D_1$ and $D_2$, and in the polarization of the photons that reach $D_3$ and $D_4$ (see panes (a-d) and (b-c)). 
In order to preserve the interference between these two paths, there must be no degree of freedom (other than polarization) which provides the which-path information. In the case of identical QDs, with zero fine-structure splitting, the which-path information is partially provided by the time correlation between the red and blue photons; hence the nonoptimal value of $P$ (Eq. \ref{PQRST}).   
In order to support such statement, we note that $ r_{kl} = q_{kl} $ in the time domain where both functions are finite, but such domain is different in the two cases (Eq. \ref{qrst}). Let's consider, for example, the case where a blue photon in D$_3$ is detected before a red one in D$_1$ ($ t_3 < t_1 $, where $ q_{kl} > r_{kl} = 0 $). Such two photons cannot proceed from the same emitter, for each dot emits two photons in a well defined time ordering, with the red photon first. This allows one to distinguish from one another the paths represented in panels (a-c) or (b-d), and the distinguishability of the two paths results in a partial loss of the interference visibility $\mathcal{V} \equiv |R|/Q $. 
In fact, the visibility is here directly related to the probability $\alpha$ that one of the two dots emits a blue photons before the other has emitted a red one \cite{SM}:
\begin{equation}
\alpha = \frac{\Gamma'}{2\Gamma+\Gamma'} = 1 - \mathcal{V} . 
\end{equation}

Besides the time correlation between the red and blue photons, the which-path information can be provided by the fine-structure splitting and by possible differences in the emission frequencies of the two dots.  
In order to account for these effects, we generalize Eq. \ref{PQRST} to the case of nonidentical QDs with finite zero-field splitting. Here, the coherence functions corresponding to probability densities are unaffected by such energy splitting, and so is the resulting term $Q$. The term related to interference takes instead the form \cite{SM}:  
\begin{eqnarray}\label{eqR1}
\frac{R}{Q} \!\!&=&\!\!
{\rm Real} \left\{ 
\frac{\Gamma'}{2\Gamma'\!+i(\omega^A_{32}-\omega^B_{32})}
\left(
\frac{\Gamma'}{\Gamma'\!+\!i\omega_{32}^A}
+
\frac{\Gamma'}{\Gamma'\!-\!i\omega_{32}^B}
\right)\right.
\nonumber\\
& & \!\!\!\left.\left[\! 
\frac{\Gamma}{2\Gamma\!\!+\!\Gamma'\!+\!i(\omega^A_{42}\!-\!\omega^B_{42})}
\!+\!
\frac{\Gamma}{2\Gamma\!\!+\!\Gamma'\!-\!i(\omega^A_{43}\!-\!\omega^B_{43})}
\!\right]\!
\right\}\! ,
\end{eqnarray}
where $\omega_{ij} \equiv E_i - E_j$.
Equation \ref{eqR1} summarizes the loss of interference visibility resulting from the different emission frequencies of the two dots, the fine-structure splitting of each dot, and the time correlation between two cascaded photons. 

{\it Partial erasure of the which-path information --}
In the subspace corresponding to two photon pairs with orthogonal polarizations, the which-path information which results from the time correlation and the fine-structure splitting can be erased. In particular, a simple combination of polarizing beam splitters and polarization rotators can implement unitary transformations of the modes $a'$ and $b'$, described by the source-field relations \cite{SM}: 
\begin{eqnarray}\label{sfr3}
a_{a'H}\! =\! \sigma_{4B} ,\ 
a_{a'V}\! =\! \sigma_{4A} ,\ 
a_{b'H}\! =\! \sigma_{3B} ,\ 
a_{b'V}\! =\! \sigma_{3A} ,
\end{eqnarray}
which replace Eq. \ref{sfr2}. As a result, the relevant interference is now between paths that differ from one another only in the source of the photon that reaches each detector. This makes the time correlation between the photons identical in the two paths, and the polarization dependence of the emission energy within each dot irrelevant. The contribution to $P$ arising from four-photon interference thus depends only on the energy difference between corresponding transitions in the two dots:
\begin{eqnarray}\label{eqR2}
\frac{R}{Q} \!\!&=&\!\!
{\rm Real} 
\left[
\!\frac{1}{2\!\!+\!i\delta_{32}}\!
\left(
\!\frac{1}{3\!+\!i\delta_{42}}\!
+
\!\frac{1}{3\!-\!i\delta_{43}}\!
\right)
\!\!
\left(
\!\frac{1}{1\!-\!i\delta_{21}}\!
+
\!\frac{1}{1\!+\!i\delta_{31}}\!
\right)
\right.
\nonumber\\
&+& 
\left.\!
\!\frac{1}{2\!+\!i\delta_{41}}\!
\left(
\!\frac{1}{3\!+\!i\delta_{42}}
\frac{1}{1\!+\!i\delta_{31}}\!
+
\!\frac{1}{3\!-\!i\delta_{43}}
\frac{1}{1\!-\!i\delta_{21}}\!
\right)
\!\right] ,
\end{eqnarray}
where $ \delta_{ij} \equiv (\omega_{ij}^A-\omega_{ij}^B) / \Gamma $ and $\Gamma' = \Gamma$. 
As is apparent from the above equation, the contribution related to interference achieves the theoretical maximum ($R=Q$) for $\delta_{ij}=0$ (Fig. \ref{fig3}(a)). Such condition corresponds to quantum emitters with identical emission frequencies, but doesn't imply a vanishing fine-structure splitting in either dot. Besides, no indistinguishability is required between two photons emitted in cascade by a given dot, and two emitted by different dots in distinct cascade decays. In the absence of a quantum erasure, the maximum of $|R|$ can only be achieved if both dots have vanishing fine-structure splitting (panel (b)), and remains below the theoretical limit unless $ \Gamma \gg \Gamma'$. 

In conclusion, we have shown that the efficiency of the entanglement swapping procedure based on the cascade decay of QDs cannot be reduced to the amount of frequency-polarization entanglement present within each pair. A relevant role is also played by the differences in the emission frequencies of the two dots and by the time correlation between cascaded photons. These effects, along with the fine-structure splitting, potentially provide the which-path information, thus reducing the visibility of the four-photon interference, and the success probability of the entanglement swapping procedure. 
The detrimental effect of time correlation can be eliminated by engineering the ratio of the photon emission rates. For identical emitters, the which-path information encoded in the fine-structure splitting can be erased by a simple manipulation of the blue-photon polarization.

Fruitful discussions with Carlos Tejedor are gratefully acknowledged.
This work was financially supported by the Italian FIRB Project No. RBFR12RPD1 of the Italian MIUR.

\clearpage

\begin{widetext}

\section{Appendix A: Derivation of the coherence functions}

Each of the two quantum dots (QDs) $A$ and $B$ is described by an Hamiltonian 
$ H_0 = \omega' (|2 \rangle\langle 2|+|3 \rangle\langle 3|) + (\omega +\omega') |4 \rangle\langle 4|$, and is intialized ($t=0$) in the highest state $|4\rangle$.
The system dynamics is driven by the radiative relaxation processes $1-4$ (see Eqs. 1-2 of the manuscript).  At $t>0$, the QD density matrix is thus given by the following mixture of the eigenstates $|k\rangle$ (with $k=1-4$):
\begin{equation}
\rho  (t) = \sum_{k=1}^4 \rho_{kk} (t) | k \rangle\langle k | = | 4 \rangle\langle 4 | e^{-2\Gamma t} +  
(| 2 \rangle\langle 2 |+| 3 \rangle\langle 3 |) \frac{\Gamma}{\Gamma'-2\Gamma} (e^{-2\Gamma t} - e^{-\Gamma' t}) + | 1 \rangle\langle 1 | \left[1-\sum_{k=2}^4 \rho_{kk} (t)\right] .
\end{equation}

The multi-time expectation values, such as the coherence functions 
$G_{ijkl} (t_1,t_3,t_4,t_2)$ can be computed by applying the quantum regression theorem. 
Let's start by considering two-time expectation values of the form 
$G_{ijkl} (t_1,t_2,t_2,t_1)$, for example with $i=l=1$ and $j=k=3$. 
The second-order coherence function is computed in the following steps:
\begin{align}
\rho (t_1=0) & = |4 \rangle\langle 4 | 
\ \xrightarrow{\mathcal{L}(t_1)}\ 
\rho (t_1)
\ \longrightarrow\ 
\rho' (t_2=t_1) = \sigma_1 \rho (t_1) \sigma_1^\dagger
              = \Gamma | 2 \rangle\langle 2 | \rho_{44} (t_1)
\nonumber\\
& 
\xrightarrow{\mathcal{L}(t_2-t_1)}
\rho' (t_2  ) = \Gamma\rho_{44} (t_1) \left[(| 2 \rangle\langle 2 |-| 1 \rangle\langle 1 | )e^{-\Gamma' (t_2-t_1)}+| 1 \rangle\langle 1 | \right]
\longrightarrow
{\rm Tr} [\sigma_3^\dagger \sigma_3 \rho' (t_2) ] = 
\Gamma' {\rm Tr} [| 2 \rangle\langle 2 | \rho' (t_2) ] .
\end{align}
Therefore, the density matrix evolves for a time $t_1$ under the effect of the superoperator $\mathcal{L}$, corresponding to the master equation in the Lindblad form (Eqs. 1-2 of the manuscript), and the density matrix $\rho'(t_2=t_1)$ is obtained by applying to $\rho (t_1)$ the ladder operators specified by the indeces $i$ and $l$. Then $\rho'$ evolves under the effect of the superoperator $\mathcal{L}$ for a time $t_2-t_1$. Finally, the coherence function corresponds to the expectation value in $\rho' (t_2)$ of the operator $\sigma^\dagger_j\sigma_k$.

The procedure is similar for a coherence function with $i \neq l$ and $j \neq k$, such as, 
for example, $G_{1342} (t_1,t_2,t_2,t_1)$
In this case, the four steps outlined above lead to the following result:
\begin{align}
\rho (t_1=0) & = |4 \rangle\langle 4 | 
\ \xrightarrow{\mathcal{L}(t_1)}\ 
\rho (t_1)
\ \longrightarrow\ 
\rho' (t_2=t_1) = \sigma_2 \rho (t_1) \sigma_1^\dagger
              = | 3 \rangle\langle 2 | \Gamma\rho_{44} (t_1)
\nonumber\\
& 
\ \xrightarrow{\mathcal{L}(t_2-t_1)}\ 
\rho' (t_2  ) = \Gamma\rho_{44} (t_1)       | 3 \rangle\langle 2 |e^{-\Gamma' (t_2-t_1)} 
\ \longrightarrow\ 
{\rm Tr} [\sigma_3^\dagger \sigma_4 \rho' (t_2) ] = 
\Gamma'{\rm Tr} [| 2 \rangle\langle 3 | \rho' (t_2) ] .
\end{align}

Depending on the indeces $i$ and $j$, the density matrix $\rho'$ can be initialized in different ways. In computing the coherence functions, we have made use of the following equations for the time evolution of its matrix elements 
$\rho_{mn}' = \langle m | \rho' | n \rangle $, 
induced by the superoperator $\mathcal{L}$:
\begin{equation}\label{eq4}
\frac{\rho_{42}' (t_2)}{\rho_{42}' (t_1)} = 
\frac{\rho_{43}' (t_2)}{\rho_{43}' (t_1)} = 
e^{-[(\Gamma+\Gamma'/2)+i\omega](t_2-t_1)}, \ 
\frac{\rho_{23}' (t_2)}{\rho_{23}' (t_1)} = 
e^{-        \Gamma'   (t_2-t_1)  } .
\end{equation}
We also note that a density matrix $\rho'$ initialized in state $|2\rangle$ or $|3\rangle$ evolves into a mixture of such state and of the ground state $|1\rangle$. Anaolgously, if $\rho'$ is initially proportional to $|2 \rangle\langle 3 | $, $|2 \rangle\langle 4 | $, or $|3 \rangle\langle 4 | $, then the same applies to $ \rho' (t_2>t_1) $. Therefore, and in view of the above equations, it is easily verified that the functions
$G_{ijkl} (t_1,t_2,t_2,t_1)$ vanish identically for $j \neq i+2$ and $k
\neq l+2$. 

We finally consider the more general case of the three-time expectation values, which can be computed by applying three times the quantum regression theorem. The required steps are illustrated hereafter for the case $i=l=1$ and $j=k=3$ (with $t_1<t_2<t_3<t_4$): 
\begin{align}
\rho (t_1=0) & = |4 \rangle\langle 4 | 
\ \xrightarrow{\mathcal{L}(t_1)}\ 
\rho (t_1)
\ \longrightarrow\ 
\rho' (t_2=t_1) = \rho (t_1) \sigma_1^\dagger
              = | 4 \rangle\langle 2 | \sqrt{\Gamma}\rho_{44} (t_1)
\xrightarrow{\mathcal{L}(t_2-t_1)}
\nonumber\\
\rho' (t_2  ) & = \sqrt{\Gamma}\rho_{44} (t_1) | 4 \rangle\langle 2 |  e^{-[(\Gamma+\Gamma'/2)+i\omega](t_2-t_1)}
\ \longrightarrow\ 
\rho'' (t_3=t_2) = \sigma_1 \rho' (t_2) 
              = | 2 \rangle\langle 2 | \sqrt{\Gamma}\rho_{42}' (t_2)
\xrightarrow{\mathcal{L}(t_3-t_2)}
\nonumber\\
&
\rho'' (t_3) = | 2 \rangle\langle 2 | \sqrt{\Gamma}\rho_{42}' (t_2) e^{-\Gamma' (t_3-t_2)} 
\ \longrightarrow\ 
\rho'''(t_4=t_3)=\rho''(t_3)\sigma_3^\dagger=| 2 \rangle\langle 1 | \sqrt{\Gamma'}\rho_{22}''(t_3)
\xrightarrow{\mathcal{L}(t_4-t_3)}
\nonumber\\
&
\ \ \ \ \ \ \ \rho''' (t_4) = \sqrt{\Gamma'}\rho_{22}'' (t_3) | 2 \rangle\langle 1 |  e^{-(\Gamma'/2+i\omega')(t_4-t_3)}
\ \longrightarrow\ 
\sqrt{\Gamma'}{\rm Tr} [\sigma_3 \rho''' (t_4) ] 
\end{align}
The calculation of the coherence functions with $i \neq l$ and $j \neq k$ proceeds along the same lines. As in the case of the two-time expectation values, one can easily show that all coherence functions with $j \neq i+2$ and $k \neq l+2$ vanish identically. 

\section{Appendix B: Derivation of the operators $X_{cd}$ and $Y_{a'b'}$}

If state at the input modes $a$ and $b$ of the beam splitter is given by the Bell state 
$ |\Psi^-_{ab}\rangle = ( | 1_{aH} , 1_{bV} \rangle - | 1_{aV} , 1_{bH} \rangle ) / \sqrt{2} $,
then the state at the output ports takes the same form: 
$ |\Psi^-_{cd}\rangle = ( | 1_{cH} , 1_{dV} \rangle - | 1_{cV} , 1_{dH} \rangle ) / \sqrt{2} $ 
(see the input output relations in Eq. 6 of the manuscript).
All the other Bell states at the input ports $a$ and $b$ evolve into states where both photons are localized in the same mode, either $c$ or $d$. The probability corresponding to $ |\Psi^-_{ab}\rangle $ can thus be identified with that of detecting a coincidence in $c$ and $d$, provided that the maximum number of photons in the system is 2.

The probability of having, in the modes $a'$ and $b'$, the Bell state 
$ |\Psi_{a'b'}^-\rangle = ( | 1_{a'H} , 1_{b'V} \rangle - | 1_{a'V} , 1_{b'H} \rangle ) / \sqrt{2} $
can be identified with the expectation value of the operator $Y_{a'b'}$ (see Eq. 9 in the manuscript). The underlying assumption is that the density matrix of the two modes $\rho_{a'b'}$ only includes states corresponding to no more than two photons.
If this is the case, then:
\begin{align}
&
{\rm Tr} \left[\rho_{a'b'} |\Psi_{a'b'}^-\rangle\langle\Psi_{a'b'}^-| \right] 
= \frac{1}{2} {\rm Tr} \left[\rho \left(
 a^\dagger_{a'H} a^\dagger_{b'V}|0 \rangle\langle 0|a_{a'H}a_{dV}
+a^\dagger_{a'V} a^\dagger_{b'H}|0 \rangle\langle 0|a_{a'V}a_{dH}
-a^\dagger_{a'H} a^\dagger_{b'V}|0 \rangle\langle 0|a_{a'V}a_{dH}
\right.\right.
\\
& \left.\left. 
-a^\dagger_{a'V} a^\dagger_{b'H}|0 \rangle\langle 0|
a_{a'H}a_{b'V} \right)\right] = \frac{1}{2} {\rm Tr} \left[\rho_{a'b'} \left(
 a^\dagger_{a'H} a^\dagger_{b'V} \mathcal{I} a_{a'H}a_{b'V}
+a^\dagger_{a'V} a^\dagger_{b'H} \mathcal{I} a_{a'V}a_{b'H}
-a^\dagger_{a'H} a^\dagger_{b'V} \mathcal{I} a_{a'V}a_{b'H}
\right.\right.
\\
& \left.\left.
-a^\dagger_{a'V} a^\dagger_{b'H} \mathcal{I} a_{a'H}a_{b'V}
\right)\right] = \langle Y_{a'b'} \rangle .
\end{align}
The second equation above is obtained by replacing the projector on the vacuum state $|0\rangle$ with the identity operator
\begin{equation}
\mathcal{I} = |0 \rangle\langle 0 | + \sum_{n_{a'V}+n_{a'H}+n_{b'V}+n_{b'H}>0} 
| n_{a'V},n_{a'H},n_{b'V},n_{b'H} \rangle\langle n_{a'V},n_{a'H},n_{b'V},n_{b'H} | ,
\end{equation}
and is satisfied because the density matrix $\rho_{a'b'}$ only includes, by assumption, only zero-, one-, and two-photon states. The expectation value of a term obtained by applying twice the creation operators to any projector of $\mathcal{I}$, other than $|0 \rangle\langle 0 |$, would thus vanish identically. 

\section{Appendix C: Derivation of the four-photon probabilities and coherences}

The probability related to the detection of four photons, one for each of the modes $c,d,a',b'$, at times $t_1, t_2, t_3, t_4$, respectively, is given by the following expression:  
\begin{subequations}
\begin{align}
p (t_1,t_2,t_3,t_4) & = \frac{\Gamma^2}{2}
\left[ \langle
a^\dagger_{cH} (t_1)\, a^\dagger_{dV} (t_2)\, a^\dagger_{a'H} (t_3)\, a^\dagger_{b'V} (t_4) 
a_{b'V}        (t_4)\, a_{a'H}        (t_3)\, a_{dV}          (t_2)\, a_{cH}          (t_1) 
\rangle \right.
\\
&+ \left. \langle
a^\dagger_{cH} (t_1)\, a^\dagger_{dV} (t_2)\, a^\dagger_{a'V} (t_3)\, a^\dagger_{b'H} (t_4) 
a_{b'H}        (t_4)\, a_{a'V}        (t_3)\, a_{dV}          (t_2)\, a_{cH}          (t_1) 
\rangle \right.
\\
&+ \left. \langle
a^\dagger_{cV} (t_1)\, a^\dagger_{dH} (t_2)\, a^\dagger_{a'H} (t_3)\, a^\dagger_{b'V} (t_4) 
a_{b'V}        (t_4)\, a_{a'H}        (t_3)\, a_{dH}          (t_2)\, a_{cV}          (t_1) 
\rangle \right.
\\
&+ \left. \langle
a^\dagger_{cV} (t_1)\, a^\dagger_{dH} (t_2)\, a^\dagger_{a'V} (t_3)\, a^\dagger_{b'H} (t_4) 
a_{b'H}        (t_4)\, a_{a'V}        (t_3)\, a_{dH}          (t_2)\, a_{cV}          (t_1) 
\rangle \right.
\\
&- \left. \langle
a^\dagger_{cH} (t_1)\, a^\dagger_{dV} (t_2)\, a^\dagger_{a'H} (t_3)\, a^\dagger_{b'V} (t_4) 
a_{b'H}        (t_4)\, a_{a'V}        (t_3)\, a_{dV}          (t_2)\, a_{cH}          (t_1) 
\rangle \right.
\\
&- \left. \langle
a^\dagger_{cH} (t_1)\, a^\dagger_{dV} (t_2)\, a^\dagger_{a'V} (t_3)\, a^\dagger_{b'H} (t_4) 
a_{b'V}        (t_4)\, a_{a'H}        (t_3)\, a_{dV}          (t_2)\, a_{cH}          (t_1) 
\rangle \right.
\\
&- \left. \langle
a^\dagger_{cV} (t_1)\, a^\dagger_{dH} (t_2)\, a^\dagger_{a'H} (t_3)\, a^\dagger_{b'V} (t_4) 
a_{b'H}        (t_4)\, a_{a'V}        (t_3)\, a_{dH}          (t_2)\, a_{cV}          (t_1) 
\rangle \right.
\\
&- \left. \langle
a^\dagger_{cV} (t_1)\, a^\dagger_{dH} (t_2)\, a^\dagger_{a'V} (t_3)\, a^\dagger_{b'H} (t_4) 
a_{b'V}        (t_4)\, a_{a'H}        (t_3)\, a_{dH}          (t_2)\, a_{cV}          (t_1) 
\rangle \right] .
\end{align}
\end{subequations}
Here, the terms (a-d) correspond to the probability of detecting photons of defined linear polarization (and defined frequency) at each of the detectors. In particular, the modes $c$ and $d$ (coupled to $D_1$ and $D_2$, see Fig. 1 of the manuscript) correspond to a frequency $\omega$, while $a'$ and $b'$ (coupled to $D_3$ and $D_4$) correspond to a frequency $\omega'$. The remaining four terms (e-h) cannot be individually interpreted in terms of a probability. They rather correspond to fourth-order coherences, and account for quantum interference effects affecting the four-photon detection. More specifically, 
the interference is between paths that differ from one another with respect to the linear polarizations of the photons at the modes $a'$ and $b'$. This can be seen by comparing, in each term, the pedices of the time-ordered creation operators with those of the time-ordered annihilation operators.

In order to compute the above expectation values, we first express the operators of the modes $c$ and $d$ as a function of those of $a$ and $b$, by means of the input-ouptut relation of the balanced beam splitter (see Eq. 6 of the manuscript). Then, the field operators are replaced by the ladder operators of the corresponding QD transitions, by applying the source-field relation (Eq. 7). Here, we assume that the traveling time of each photon from the source to the detector is the same, such that the common time delay between the photon emission and its detection can be omitted. The resulting expression of the probability density is given by:   
\begin{subequations}\label{eq2}
\begin{align}
p (t_1,t_2,t_3,t_4) & = \frac{1}{8}
\left[
\langle
\sigma_{1A}^\dagger (t_1) \,\sigma_{2B}^\dagger (t_2) \,\sigma_{3A}^\dagger (t_3) \,\sigma_{4B}^\dagger (t_4) 
\sigma_{4B}         (t_4) \,\sigma_{3A}         (t_3) \,\sigma_{2B}         (t_2) \,\sigma_{1A}         (t_1) 
\rangle \right.
\\
&+ \left.
\langle
\sigma_{2A}^\dagger (t_1) \,\sigma_{1B}^\dagger (t_2) \,\sigma_{4A}^\dagger (t_3) \,\sigma_{3B}^\dagger (t_4) 
\sigma_{3B}         (t_4) \,\sigma_{4A}         (t_3) \,\sigma_{1B}         (t_2) \,\sigma_{2A}         (t_1) 
\rangle \right.
\\
&+ \left.
\langle
\sigma_{1B}^\dagger (t_1) \,\sigma_{2A}^\dagger (t_2) \,\sigma_{3B}^\dagger (t_3) \,\sigma_{4A}^\dagger (t_4) 
\sigma_{4A}         (t_4) \,\sigma_{3B}         (t_3) \,\sigma_{2A}         (t_2) \,\sigma_{1B}         (t_1) 
\rangle \right.
\\
&+ \left.
\langle
\sigma_{2B}^\dagger (t_1) \,\sigma_{1A}^\dagger (t_2) \,\sigma_{4B}^\dagger (t_3) \,\sigma_{3A}^\dagger (t_4) 
\sigma_{3A}         (t_4) \,\sigma_{4B}         (t_3) \,\sigma_{1A}         (t_2) \,\sigma_{2B}         (t_1) 
\rangle \right.
\\
&+ \left.
\langle
\sigma_{1A}^\dagger (t_1) \,\sigma_{2B}^\dagger (t_2) \,\sigma_{3A}^\dagger (t_3) \,\sigma_{4B}^\dagger (t_4) 
\sigma_{3B}         (t_4) \,\sigma_{4A}         (t_3) \,\sigma_{2A}         (t_2) \,\sigma_{1B}         (t_1) 
\rangle \right.
\\
&+ \left.
\langle
\sigma_{2A}^\dagger (t_1) \,\sigma_{1B}^\dagger (t_2) \,\sigma_{4A}^\dagger (t_3) \,\sigma_{3B}^\dagger (t_4) 
\sigma_{4B}         (t_4) \,\sigma_{3A}         (t_3) \,\sigma_{1A}         (t_2) \,\sigma_{2B}         (t_1) 
\rangle \right.
\\
&+ \left.
\langle
\sigma_{1B}^\dagger (t_1) \,\sigma_{2A}^\dagger (t_2) \,\sigma_{3B}^\dagger (t_3) \,\sigma_{4A}^\dagger (t_4) 
\sigma_{3A}         (t_4) \,\sigma_{4B}         (t_3) \,\sigma_{2B}         (t_2) \,\sigma_{1A}         (t_1) 
\rangle \right.
\\
&+ \left.
\langle
\sigma_{2B}^\dagger (t_1) \,\sigma_{1A}^\dagger (t_2) \,\sigma_{4B}^\dagger (t_3) \,\sigma_{3A}^\dagger (t_4) 
\sigma_{4A}         (t_4) \,\sigma_{3B}         (t_3) \,\sigma_{1B}         (t_2) \,\sigma_{2A}         (t_1) 
\rangle
\right] .
\end{align}
\end{subequations}
Also these terms can be divided into two groups. In the first four (lines (a-d)), the pedices of the time-ordered raising operators coincide with those of the time-ordered lowering operators. These terms are real and nonnegative, and each of them can be interpreted as a probability and associated with one of the four ways in which the four emitted photons can reach the four detectors (Fig. 2 of the manuscript). The remaining four terms (e-h) can be interpreted in terms of quantum interference. In particular, the interference is between paths characterized by the same polarizations but different origin (dot $A$ or $B$) of the red photons that reach $D_1$ (at time $t_1$) and $D_2$ (at time $t_2$), and same source but different polarization of the photons that reach $D_3$ and $D_4$.

Being the dynamics of each dot independent from that of the other one, each of the above expectation values can be factorized into the product of two second-order coherence functions, one for each QD:
\begin{subequations}\label{eq1}
\begin{align}
p & = \frac{1}{8}
\left[
G^A_{1331} (t_1,t_3,t_3,t_1) \,
G^B_{2442} (t_2,t_4,t_4,t_2)
+\ 
G^A_{2442} (t_1,t_3,t_3,t_1) \,
G^B_{1331} (t_2,t_4,t_4,t_2)
\right.
\\
&+ \left.
G^B_{1331} (t_1,t_3,t_3,t_1) \,
G^A_{2442} (t_2,t_4,t_4,t_2)
+\ 
G^B_{2442} (t_1,t_3,t_3,t_1) \,
G^A_{1331} (t_2,t_4,t_4,t_2)
\right.
\\
&+ \left.
G^A_{1342} (t_1,t_3,t_3,t_2) \,
G^B_{2431} (t_2,t_4,t_4,t_1)
+\ 
G^A_{2431} (t_1,t_3,t_3,t_2) \,
G^B_{1342} (t_2,t_4,t_4,t_1)
\right.
\\
&+ \left.
G^B_{1342} (t_1,t_3,t_3,t_2) \,
G^A_{2431} (t_2,t_4,t_4,t_1)
+\ 
G^B_{2431} (t_1,t_3,t_3,t_2) \,
G^A_{1342} (t_2,t_4,t_4,t_1)
\right] .
\end{align}
\end{subequations}

By replacing in Eq. \ref{eq1} the expressions of the second-order coherence functions (Eq. 4 of the manuscript), one can show that all 8 contributions are finite and identical to one another for $t_1,t_2<t_3,t_4$. For $t_1>t_3$ and $t_2<t_4$, or $t_1<t_3$ and $t_2>t_4$, instead, the former 4 terms (lines (a-d)) are still finite, whereas the latter four contributions (e-h) vanish identically (Eq. 15 of the manuscript). 

\section{Appendix D: Derivation of the probability $\alpha$}

The probability that one QD emits a blue photon before the other dot emits a red photon is given by the following expression:
\begin{equation}
\alpha = \sum_{i=1}^2 \sum_{\chi = A,B} \Gamma \int_0^\infty dt \rho^\chi_{44} (t) \int_0^t dt_1 \int_{t_1}^{t} dt_2 G^{\bar\chi}_{ijji} (t_1,t_2,t_2,t_1) = \frac{\Gamma'}{2\Gamma+\Gamma'} ,
\end{equation}
where $\bar A= B$, $\bar B= A$, and $j=i+2$. In fact, the probability density corresponding to the emission at time $t$ of a photon with a given polarization by a given dot $\chi$ is given by 
$ \Gamma \rho_{44}^\chi (t) $. The probability that the other dot $\bar\chi$ has already emitted two photons of, e.g., polarization $H$ at that time, is given by the time integral of
$ G_{1331} (t_1,t_2,t_2,t_1) $ over $t_1 < t_2 < t$. Being the two probabilities related to different and independent system, the joint probability is given by the product of the two above probabilties, integrated in time ($t$) from $0$ to $\infty$.
Finally, in order to obtain the overall probability $\alpha$, we sum over all the possible polarizations of the three photons, and both the possibile identities of the dot that emits the last ($\chi$).

\section{Appendix E: Derivation of the coherence functions in the presence of fine-structure splitting and spectral inhomogeneity}

If the levels 2 and 3 of each dot are not degenerate, the expressions of the second-order coherence functions considered so far have to be generalized. In particular, by following the steps outlined in the first Section of the present Supplemental Material, one can show that the functions $ G_{ijkl} (t_1,t_2,t_2,t_1) $ with $i=l$ and $j=k$ are left unaffected. Instead, the functions with $i \neq $ and $j \neq k$ become ($t_1<t_2<t_3<t_4$):
\begin{eqnarray}
G_{1342} (t_1,t_3,t_4,t_2) & = &  
\Gamma \Gamma' 
e^{-2\Gamma t_1 - [(\Gamma+\Gamma'/2)+i\omega_{42}](t_2-t_1)-(\Gamma'+i\omega_{32})(t_3-t_2)-(\Gamma'/2-i\omega_{21})(t_4-t_3)}.
\end{eqnarray}
The additional oscillating terms that appear in the above two equations as a consequence of the fine-structure splitting result from the modified time evolution of the coherences, which replaces the one reported in Eq. \ref{eq4}:
\begin{equation}
\rho_{32}' (t_{k+1}) = \rho_{32}' (t_k) e^{-(\Gamma'+i\omega_{32}) (t_{k+1}-t_k)  } .
\end{equation}
Besides, the frequency $\omega$ is replaced either by $\omega_{42}$ or $\omega_{43}$, depending on whether the photon polarization is $H$ or $V$, and $\omega'$ is replaced either by $\omega_{21}$ or $\omega_{31}$.

The above expression of the second-order coherence functions are used to derive the probability density  $p$. In particular, this applies to the terms $r_{kl}$ (Eq. 14 of the manuscript), whereas the $q_{kl}$ (Eq. 13) are left unchanged with respect to the case of identical dots with no fine-structure splitting. As in that case, the generalized expressions of $r_{kl}$ are summed with respect to $k$ and $l$, and time integrated, which leads to Eq. 18 of the manuscript.

\section{Appendix F: Derivation of $R$ with the partial quantum erasure}

In order to erase both the which-way information encoded in the time correlation of the cascaded photons and in the fine-structure splitting, one needs to exchange the blue photon emitted with polarization, e.g., $H$ by dot $A$ with the blue photon emitted with polarization $V$ by dot $B$. This can be achieved by rotating the linear polarization of $a'$ by $\pi /2$, then applying a polarizaing beam-splitter, and finally rotating the linear polarization back: 
\begin{eqnarray}
\sigma_{3A} & \longrightarrow & a_{a_1 V} \ \ \longrightarrow \ \ a_{b_2 V} \ \ \longrightarrow \ \ a_{b' V}
\\
\sigma_{4A} & \longrightarrow & a_{a_1 H} \ \ \longrightarrow \ \ a_{a_2 H} \ \ \longrightarrow \ \ a_{a' V} 
\\
\sigma_{3B} & \longrightarrow & a_{b_1 H} \ \ \longrightarrow \ \ a_{b_2 H} \ \ \longrightarrow \ \ a_{b' H}
\\
\sigma_{4B} & \longrightarrow & a_{b_1 V} \ \ \longrightarrow \ \ a_{a_2 V} \ \ \longrightarrow \ \ a_{a' H} .
\end{eqnarray}
The resulting change in the source-field relations for the red photons (Eq. 19 of the manuscript) leads to the replacement of Eq. \ref{eq2} above with the following equation:
\begin{subequations}
\begin{align}
p (t_1,t_2,t_3,t_4) & = \frac{1}{8}
\left[
\langle
\sigma_{1A}^\dagger (t_1) \,\sigma_{2B}^\dagger (t_2) \,\sigma_{4B}^\dagger (t_3) \,\sigma_{3A}^\dagger (t_4) 
\sigma_{3A}         (t_4) \,\sigma_{4B}         (t_3) \,\sigma_{2B}         (t_2) \,\sigma_{1A}         (t_1) 
\rangle \right.
\\
&+ \left.
\langle
\sigma_{2A}^\dagger (t_1) \,\sigma_{1B}^\dagger (t_2) \,\sigma_{4A}^\dagger (t_3) \,\sigma_{3B}^\dagger (t_4) 
\sigma_{3B}         (t_4) \,\sigma_{4A}         (t_3) \,\sigma_{1B}         (t_2) \,\sigma_{2A}         (t_1) 
\rangle \right.
\\
&+ \left.
\langle
\sigma_{1B}^\dagger (t_1) \,\sigma_{2A}^\dagger (t_2) \,\sigma_{3B}^\dagger (t_3) \,\sigma_{4A}^\dagger (t_4) 
\sigma_{4A}         (t_4) \,\sigma_{3B}         (t_3) \,\sigma_{2A}         (t_2) \,\sigma_{1B}         (t_1) 
\rangle \right.
\\
&+ \left.
\langle
\sigma_{2B}^\dagger (t_1) \,\sigma_{1A}^\dagger (t_2) \,\sigma_{3A}^\dagger (t_3) \,\sigma_{4B}^\dagger (t_4) 
\sigma_{4B}         (t_4) \,\sigma_{3A}         (t_3) \,\sigma_{1A}         (t_2) \,\sigma_{2B}         (t_1) 
\rangle \right.
\\
&+ \left.
\langle
\sigma_{1A}^\dagger (t_1) \,\sigma_{2B}^\dagger (t_2) \,\sigma_{4B}^\dagger (t_3) \,\sigma_{3A}^\dagger (t_4) 
\sigma_{3B}         (t_4) \,\sigma_{4A}         (t_3) \,\sigma_{2A}         (t_2) \,\sigma_{1B}         (t_1) 
\rangle \right.
\\
&+ \left.
\langle
\sigma_{2A}^\dagger (t_1) \,\sigma_{1B}^\dagger (t_2) \,\sigma_{4A}^\dagger (t_3) \,\sigma_{3B}^\dagger (t_4) 
\sigma_{3A}         (t_4) \,\sigma_{4B}         (t_3) \,\sigma_{1A}         (t_2) \,\sigma_{2B}         (t_1) 
\rangle \right.
\\
&+ \left.
\langle
\sigma_{1B}^\dagger (t_1) \,\sigma_{2A}^\dagger (t_2) \,\sigma_{3B}^\dagger (t_3) \,\sigma_{4A}^\dagger (t_4) 
\sigma_{4B}         (t_4) \,\sigma_{3A}         (t_3) \,\sigma_{2B}         (t_2) \,\sigma_{1A}         (t_1) 
\rangle \right.
\\
&+ \left.
\langle
\sigma_{2B}^\dagger (t_1) \,\sigma_{1A}^\dagger (t_2) \,\sigma_{3A}^\dagger (t_3) \,\sigma_{4B}^\dagger (t_4) 
\sigma_{4A}         (t_4) \,\sigma_{3B}         (t_3) \,\sigma_{1B}         (t_2) \,\sigma_{2A}         (t_1) 
\rangle
\right] .
\end{align}
\end{subequations}
Once again, being the dynamics of each dot independent from that of the other one, each of the above expectation values can be factorized into the product of two second-order coherence functions:
\begin{subequations}
\begin{align}
p & = \frac{1}{8}
\left[
G^A_{1331} (t_1,t_4,t_4,t_1) \,
G^B_{2442} (t_2,t_3,t_3,t_2)
+\ 
G^A_{2442} (t_1,t_3,t_3,t_1) \,
G^B_{1331} (t_2,t_4,t_4,t_2)
\right.
\\
&+ \left.
G^B_{1331} (t_1,t_3,t_3,t_1) \,
G^A_{2442} (t_2,t_4,t_4,t_2)
+\ 
G^B_{2442} (t_1,t_4,t_4,t_1) \,
G^A_{1331} (t_2,t_3,t_3,t_2)
\right.
\\
&+ \left.
G^A_{1342} (t_1,t_4,t_3,t_2) \,
G^B_{2431} (t_2,t_3,t_4,t_1)
+\ 
G^A_{2431} (t_1,t_3,t_4,t_2) \,
G^B_{1342} (t_2,t_4,t_3,t_1)
\right.
\\
&+ \left.
G^B_{1342} (t_1,t_3,t_4,t_2) \,
G^A_{2431} (t_2,t_4,t_3,t_1)
+\ 
G^B_{2431} (t_1,t_4,t_3,t_2) \,
G^A_{1342} (t_2,t_3,t_4,t_1)
\right] .
\end{align}
\end{subequations}
Integration in time of the probability density $p$ finally leads to:
\begin{subequations}
\begin{align}
\frac{R}{Q} & = 
{\rm Real} 
\left\{
\frac{\Gamma'}{2\Gamma'\!+\!i\delta_{32}}
\left[
\left(
\frac{\Gamma}{2\Gamma\!+\!\Gamma'\!+\!i\delta_{42}}
+
\frac{\Gamma}{2\Gamma\!+\!\Gamma'\!-\!i\delta_{43}}
\right)
\left(
\frac{\Gamma'}{\Gamma'\!-\!i\delta_{21}}
+
\frac{\Gamma'}{\Gamma'\!+\!i\delta_{31}}
\right)
\right]
\right.
\nonumber\\
& \left. +
\frac{\Gamma'}{2\Gamma'\!+\!i\delta_{41}}
\left(
\frac{\Gamma}{2\Gamma\!+\!\Gamma'\!+\!i\delta_{42}}
\frac{\Gamma'}{\Gamma'\!+\!i\delta_{31}}
+
\frac{\Gamma}{2\Gamma\!+\!\Gamma'\!-\!i\delta_{43}}
\frac{\Gamma'}{\Gamma'\!-\!i\delta_{21}}
\right)
\right\} ,
\end{align}
\end{subequations}
which reduces to Eq. 20 of the manuscript for $\Gamma = \Gamma'$.

\end{widetext}

\end{document}